\documentclass[12pt]{conm-p-l}
\usepackage{amsmath,amssymb,amsthm,latexsym, times, mathptm,graphicx}
\newtheorem{theorem}{Theorem}
\newtheorem{lemma}[theorem]{Lemma}

\def\CC{\Bbb{C}}
\def\RR{\Bbb{R}}
\def\pf{{\bf Proof }}

\begin{document}
\title{On the limiting absorption principle and spectra of quantum graphs}
\author{Beng-Seong Ong}
\address{Department of Mathematics, Texas A\& M University, College Station, TX 77843-3368}

\email{bsong@math.tamu.edu}
\date{September 28, 2005}

\maketitle

\begin{abstract}
The main result of the article is validity of the limiting
absorption principle and thus absence of the singular continuous
spectrum for compact quantum graphs with several infinite leads
attached. The technique used involves Dirichlet-to-Neumann
operators.
\end{abstract}

\section{Introduction}

The object of study in this paper is a quantum graph $\Gamma$. The
reader can find surveys of main definitions, properties, and
origins of quantum graphs, as well as main references in
\cite{ExS3,KoS1,KoS2,Ku_isaac,Ku02,Ku04_graphs1,QGraphs}. We will
briefly summarize the notions that we will need here.

Consider a compact graph $\Gamma_0$, whose edges are equipped with
coordinates (called $x$, or $x_e$, if we need to specify the edge
$e$) that identify them with segments of the real axis\footnote{We
will call the corresponding coordinate $x$, usually without
specifying the edge, which should not lead to a confusion.}.  A
finite set $B$ of vertices of cardinality $|B|=n$, which we will
call the {\em boundary of} $\Gamma_0$ is assumed to be fixed. Each
vertex $v\in B$ has an infinite edge ({\em lead}) $e_v$ attached,
which is equipped with a coordinate that identifies it with the
positive semi-axis. Thus an infinite graph $\Gamma$ is formed (see
Fig. 1).
\begin{figure}[ht]
\begin{center}
\scalebox{.5}{\includegraphics{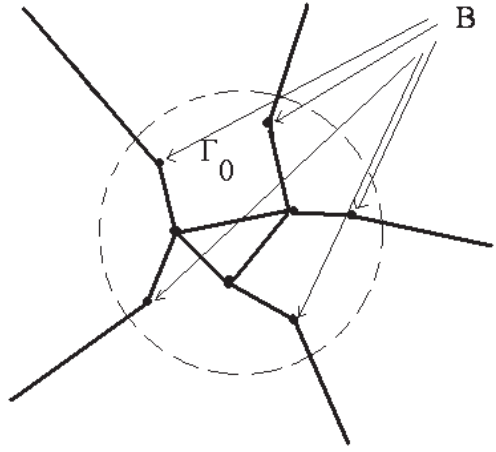}}
\end{center}
\caption{Graph $\Gamma$.}
\end{figure}
One can naturally define differentiation of functions on $\Gamma$
along the edges, as well as integration over $\Gamma$. In
particular, the space $L_2(\Gamma)$ can be defined, as well as
spaces Sobolev spaces $H^s(e)$ of functions on any edge $e$.

This graph is turned into a {\em quantum graph} by defining on it
a self-adjoint differential operator $H$ as follows:\\
The operator $H$ acts on functions on $\Gamma$ as the negative
second derivative $-\frac{d^2}{dx^2}$ along the edges. Its domain
consists of functions $f$ that belong to the Sobolev space
$H^2(e)$ on each edge $e$ of $\Gamma$ and satisfy the following
boundary conditions at the vertices:
\begin{equation}\label{E:vertex_condit}
A_v {\mathbf f}(v) + B_v {\mathbf f^\prime}(v)=0\,\,\mbox{for each
vertex } v.
\end{equation}
Here ${\mathbf f}(v)$ and ${\mathbf f^\prime}(v)$ are
correspondingly the vector of values of $f$ at $v$ attained from
directions of different edges converging at $v$ and the vector of
derivatives at $v$ in the outgoing directions. Matrices $A_v$ and
$B_v$ are of size $d_v\times d_v$, where $d_v$ is the degree of
$v$ and satisfy the following two conditions:
\begin{equation}\label{E:matrix_condit}
\begin{cases}
  rank(A_v\,\,\, B_v)=d_v \\
  A_v\times B_v^* \,\, \mbox{ is self-adjoint}.\\
\end{cases}
\end{equation}
It is known that (\ref{E:vertex_condit})-(\ref{E:matrix_condit})
represent the most general (local) self-adjoint boundary value
conditions for the operator we consider, see \cite{KoS1,
Ku04_graphs1} \footnote{One can describe in a similar way
non-local self-adjoint boundary conditions, then the matrices
should act on the vectors of values and derivatives assembled over
all graph \cite{KoS1}. The results of this paper hold for this
more general situation with no change in statements or proofs.}.

We will assume that all ``boundary'' vertices $v\in B$ have degree
two and that the boundary conditions at any such $v$ are
``Neumann'':
\begin{equation}\label{E:condit_at_boundary}
    \begin{array}{cc}
  i) & f \mbox{ is continuous at each vertex }v\in B\\
  ii) &  \mbox{at each vertex }v\in B,\,\sum\limits_{v\in e}\frac{df}{dx_e}(v)=0. \\
\end{array}.
\end{equation}
Here $\sum\limits_{v\in e}\frac{df}{dx_e}(v)$ is the sum of the
derivatives of $f$ in the outgoing directions along all edges
emanating from $v$. In fact, these conditions simply mean that the
function and its derivative are both continuous at $v$.

This assumption on the degrees of boundary vertices and on how the
conditions look like at the boundary $B$, in fact does not reduce
the generality. Indeed, one can always introduce ``fake'' boundary
vertices a little bit further away along the infinite edges and
consider them as the new boundary. Then our assumptions are
automatically satisfied, and the operator does not change at all.

It is well known (e.g., \cite{KoS1,Ku04_graphs1} and references
therein), that such defined, the operator $H$ is self-adjoint,
bounded below, and in the case of a compact graph (which $\Gamma$
is not, but $\Gamma_0$ is) has compact resolvent and thus discrete
spectrum. The structure of the spectrum of the operator $H$ on
graphs $\Gamma$ of the type described above has been studied for
quite a while (e.g.,
\cite{ExS3,GP,KoS1,KoS2,KoS3,KS,KS2,KS3,MP,Pavlov_QNetw,PF,Pavlov_DtN,QGraphs}).
It is a common knowledge that it possesses continuous part filling
the nonnegative half-axis, as well as possibly point spectrum
consisting of isolated eigenvalues accumulating to infinity. For
completeness, we provide a proof of the following standard
statement here (a variation of this proof would use Krein's
resolvent formula) that claims that the continuous part of the
spectrum does not depend on a compact part of the graph.

\begin{lemma}\label{L:continuous_spectrum} In the presence of infinite leads (i.e., if
$n>0$), the continuous spectrum of $H$ coincides with the
nonnegative half-axis. Eigenvalues of finite multiplicity
(including those embedded into the continuous spectrum)
accumulating to infinity might be present.
\end{lemma}
The proof of the lemma uses Glazman's splitting technique
\cite{Glazman}. Let us choose coordinates on each of the infinite
leads $e_v,\,v\in B$ that identify the leads with the nonnegative
half axis. We identify the point $v$ with the coordinate $x=0$. So
we have $\Gamma = \Gamma_0 \cup \Gamma_1$ where $\Gamma_1$ is the
disjoint union of $n$ copies of half-axes $[0,\infty)$. Consider
the symmetric operator $Q$ that is the restriction of $H$ on the
set of those functions $f\in D(H)$ that vanish with their first
derivative at all points $v$. Then $Q$ naturally splits into the
orthogonal sum of two minimal operators $Q_0\bigoplus Q_1$ defined
on $\Gamma_0$ and $\Gamma_1$ respectively. Since $Q_0$ acts on a
compact graph $\Gamma_0$ and $Q_1$ is just the direct sum of $n$
copies of minimal operators corresponding to $-\frac{d^2}{dx^2}$
on the half-axis, we conclude that the continuous spectrum of the
closure of $Q$ is the same as that of $Q_1$. Noticing that $H$ is
a finite dimensional extension of $Q$, one can employ Theorems 4
and 11 from \cite[Ch. I]{Glazman} to imply the statements of the
lemma.

The goal of this paper is to prove a limiting absorption principle
and thus absence of the singular continuous spectrum.
\begin{theorem}\label{T:main}
Let $R(\lambda)$ be the resolvent of $H$ and $f$ be any function
from the domain of $H$ that is compactly supported and smooth on
each edge. Then the function $(R(\lambda)f,f)$ can be analytically
continued from the upper half-plane through $\RR^+$, except for a
discrete subset of $\RR^+$. In particular, the singular continuous
spectrum $\sigma_{sc}(H)$ of $H$ is empty and the absolute
continuous spectrum coincides with the nonnegative half axis.
\end{theorem}

\section{Dirichlet-to-Neumann map and other auxiliary considerations}

Let us consider the compact part $\Gamma_0$ of our graph $\Gamma$
and treat the vertex set $B$ as its ``boundary.'' We need to
define some auxiliary objects related to $\Gamma_0$.

First of all, we will consider the operator $H_0$ on
$L_2(\Gamma_0)$ that acts as the negative second derivative along
each edge, and whose domain $D(H_0)$ consists of those functions
from the Sobolev space $H^2(e)$ on each edge $e$ of $\Gamma_0$
that are equal to zero on $B$ and satisfy conditions
(\ref{E:vertex_condit}) at all vertices of $\Gamma_0$ except those
in $B$. It is standard \cite{Ku04_graphs1} that this operator is
self-adjoint, bounded below, and has compact resolvent, and thus
discrete spectrum $\sigma(H_0)=\{\lambda_1,...,\lambda_n,...\}$
accumulating to infinity. We will denote by $R_0(\lambda)$ the
resolvent of this operator.

One can also define a linear extension operator $E$ acting from
functions defined on the (finite) set $B$ into the domain of $H_0$
$$
E: \CC^{n} \mapsto D(H_0),
$$
such that $(Ef)(v)=f(v)$ for all $v\in B$ and the derivative of
$Ef$ at each $v\in B$ along any edge of $\Gamma_0$ entering $v$ is
equal to zero\footnote{We remind the reader that $n=|B|$.}. Such
an operator is indeed easy to construct. For instance, let for any
$v\in B$ one defines a function $g_v$ such that it is equal to $1$
in a neighborhood of $v$, is smooth on each edge entering $v$, and
is supported inside the ball of radius $l_0/2$, where $l_0$ is the
smallest length of an edge of $\Gamma_0$. Then one can define
$Ef(x)=\sum\limits_{v\in B}f(v)g_v(x)$.

Another operator $N$ that we need is an analog of the ``normal
derivative at the boundary $B$ of $\Gamma_0$.'' It acts as
follows: for any function $f$ on $\Gamma_0$ that belongs to
$H^2(e)$ on any edge $e$, one can define the value $Nf(v)$ for
$v\in B$ as the derivative of $f$ at $v$ (taken in the direction
towards $v$):
$$
Nf(v)=\frac{df}{dx_e}(v),
$$
where $x_e$ is the coordinate along $e$ that increases towards $v$
We remind the reader that each vertex in $B$ has only degree two
and only one of the edges connected to each vertex in $B$ belongs
to $\Gamma_0$. Hence there is no ambiguity in defining $N$ as
above.

The main technical tool that we will use is the so called
Dirichlet-to-Neumann operator, very popular in inverse problems
\cite{Pavlov_DtN, Syl_Uhl, Uhl}, spectral theory
\cite{Pavlov_periodic, Fried2}, and since recently in quantum
graph theory \cite{Fulling, Pavlov_periodic, Ku05_jpa} as well. It
is a linear operator $\Lambda(\lambda)$ (in our case
finite-dimensional) acting on functions defined on $B$, i.e.
$\Lambda(\lambda):\CC^n\mapsto \CC^n.$ It is defined as follows.
Given a function $\phi$ on $B$, one solves the following problem
on $\Gamma_0$:
\begin{equation}\label{E:interior_problem}
 \begin{cases}
  -\frac{d^2u}{dx^2}-\lambda u=0 \mbox{ on }\Gamma_0  \\
  \mbox{conditions (\ref{E:vertex_condit}) are satisfied at all vertices of } \Gamma_0 \mbox{ except those in }B \\
  u|_B=\phi  \\
\end{cases}
\end{equation}
One now defines the {\em Dirichlet-to-Neumann map} as follows:
\begin{equation}\label{E:DtN}
    \Lambda(\lambda)\phi=N u,
\end{equation}
which justifies the name of the operator. The validity of this
definition depends upon (unique) solvability of the problem
(\ref{E:interior_problem}), which holds unless $\lambda\in
\sigma(H_0)$. Thus, $\Lambda(\lambda)$ is defined unless
$\lambda\in \sigma(H_0)$.
\begin{lemma}\label{L:DtN}
\begin{enumerate}
\item The following operator relation holds:
\begin{equation}\label{E:DtN_resolvent_relation}
\Lambda(\lambda)=NR_0(\lambda)(\frac{d^2}{dx^2}+\lambda)E
\end{equation}

\item The Dirichlet-to-Neumann map $\Lambda(\lambda)$ is a
meromorphic matrix valued function of $\lambda$ with poles on the
spectrum of $H_0$.

\item For real values $\lambda\in \RR - \sigma(H_0)$ the matrix
$\Lambda(\lambda)$ is Hermitian.
\end{enumerate}
\end{lemma}

\pf  of the Lemma. Let us introduce a new function $g=u-E\phi$ on
$\Gamma_0$. By the construction of the extension operator $E$, $g$
clearly satisfies the same vertex conditions
(\ref{E:vertex_condit}) on $\Gamma_0 - B$, as well as the zero
Dirichlet conditions on the boundary $g|_B=0$. We also note that
$Ng=Nu$, since $NE\phi=0$ for any $\phi$. This means that
(\ref{E:interior_problem}) can be equivalently rewritten as
$$
\begin{cases}
  -\frac{d^2g}{dx^2}-\lambda g=\left(\frac{d^2}{dx^2}+\lambda\right) E\phi \in L^2(\Gamma_0) \mbox{ on }\Gamma_0  \\
  \mbox{conditions (\ref{E:vertex_condit}) are satisfied at all vertices of } \Gamma_0 \mbox{ except those in }B \\
  g|_B=0  \\
\end{cases}.
$$
In other words,
$(H_0-\lambda)g=\left(\frac{d^2}{dx^2}+\lambda\right) E\phi$,
which together with the definition of the Dirichlet-to-Neumann map
proves the first statement of the Lemma.

The second statement of the lemma immediately follows from the
first one together with the discreteness of the spectrum of $H_0$
and standard analyticity properties of the resolvent.

The third statement is well known (e.g., \cite{Pavlov_periodic})
and can be checked by straightforward calculation. \qed

\section{Proof of the main result}

The proof of Theorem \ref{T:main} will use the
Dirichlet-to-Neumann map to rewrite the spectral problem on
$\Gamma$ as a vector valued spectral problem on half-line with a
general Robin condition at the origin.

First of all, Lemma \ref{L:continuous_spectrum} implies that it is
sufficient to prove absence of singular continuous spectrum on the
positive half-axis only. Then the statement about absolute
continuous spectrum would follow as well by the same Lemma.

Let $R(\lambda)$ be the resolvent of $H$. The first statement of
Theorem \ref{T:main} is established in the following
\begin{lemma}\label{L:lim_abs}
Let $f$ be a compactly supported function $\Gamma$ which is smooth
on each edge and satisfies the vertex conditions
(\ref{E:vertex_condit}). Then for any interval $[a,b]\subset
\RR^+$ that does not intersect $\sigma(H_0)$ one has
\begin{equation}\label{E:lim_abs}
    \mathop{\sup}\limits_{\begin{array}{c}
      a\leq \lambda \leq b \\
      0< \epsilon <1 \\
    \end{array}}
    |(R(\lambda+i\epsilon)f,f)|<\infty.
\end{equation}
In fact, the expression $(R(\lambda)f,f)$ can be analytically
continued through such intervals $[a,b]$.
\end{lemma}

So, now our task is to prove Lemma \ref{L:lim_abs}. This will be
done using the Dirichlet-to-Neumann operator to reduce the
spectral problem for $H$ on $\Gamma$ to a vector one on the
half-line.

At this point it will be beneficial to have in mind a different
geometric picture of $\Gamma$ than in Fig. 1. Namely, imagine that
all the $n$ infinite rays $e_v,v\in B$ are stretched along the
positive half-axis in parallel, being connected at the origin by
the finite graph $\Gamma_0$ attached to the rays at the vertices
of $B$ (see Fig. 2).
\begin{figure}[ht]
\begin{center}
\scalebox{.5}{\includegraphics{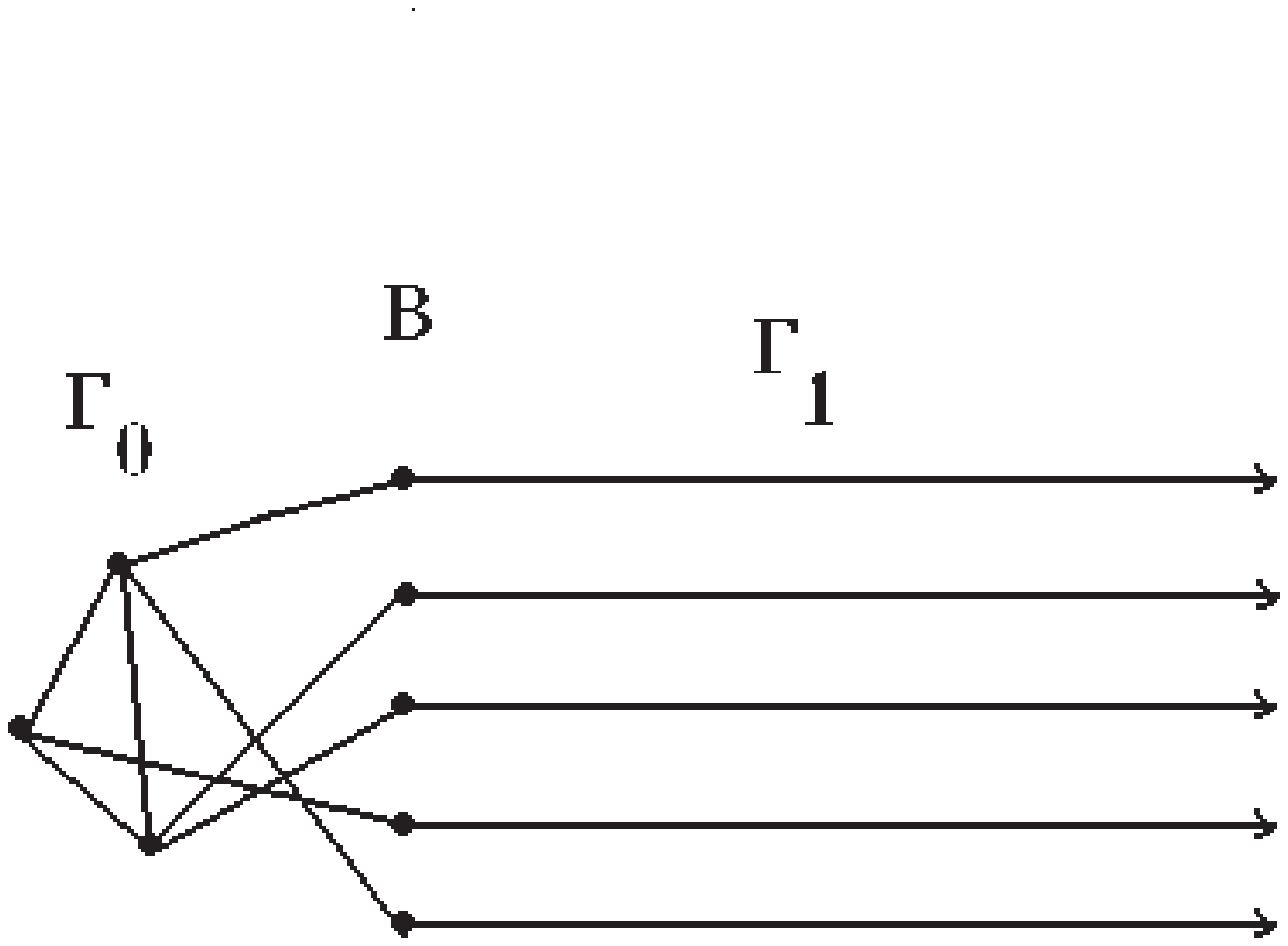}}
\end{center}
\caption{A different visualization of $\Gamma$.}
\end{figure}
Any function $u$ on $\Gamma$ can now be viewed as the pair
$(u_0,u_1)$, where $u_j=u|_{\Gamma_j}$. Functions defined on the
part $\Gamma_1$ of $\Gamma$ (in particular, $u_1$) can be
interpreted as vector-valued functions on $\RR^+$ with values in
$\CC^n$ (recall that $n=|B|$). In particular, interpreting $u_1$
as such, we can write $u|_B=u_1|_B=u(0)$, where $0$ is the origin
in $\RR^+$.

Let now $f=(f_0,f_1)$ be as in Lemma \ref{L:lim_abs}. Then
$u=R(\lambda)f$ is a function that belongs to $H^2_{loc}$ on each
edge and satisfies vertex conditions (\ref{E:vertex_condit}) and
the equation
\begin{equation}\label{E:spectral_problem}
Hu-\lambda u=f.
\end{equation}
Here $u$ naturally depends on $\lambda$. The quantity we need to
estimate in (\ref{E:lim_abs}) is now the inner product
$(u,f)=(u_0,f_0)+(u_1,f_1)$. Let us write
(\ref{E:spectral_problem}) and the vertex conditions separately
for $u_0$ on $\Gamma_0$ and $u_1$ on $\Gamma_1$. On the compact
graph $\Gamma_0$ we get
\begin{equation}\label{E:spectral_on_gamma0}
    \begin{cases}
    (-\frac{d^2}{dx^2}-\lambda)u_0=f_0\\
    \mbox{(\ref{E:vertex_condit}) satisfied on vertices of
    }\Gamma_0\mbox{ except those in }B\\
    u_0|_B=u_1(0)
    \end{cases}
\end{equation}
Similarly, on $\Gamma_1$ we have
\begin{equation}\label{E:spectral_on_gamma1}
    \begin{cases}
    (-\frac{d^2}{dx^2}-\lambda)u_1=f_1 \mbox{ on }\RR^+\\
    u^\prime_1(0)=Nu_0
    \end{cases}
\end{equation}
Here $N$ is the introduced before ``normal derivative at $B$''
operator on $\Gamma_0$ and functions $u_1,f_1$ are interpreted as
functions on $\RR^+$ with values in $\CC^n$.

Notice that the boundary conditions on $B$ in
(\ref{E:spectral_on_gamma0}) and at zero in
(\ref{E:spectral_on_gamma1}) are just the vertex conditions
(\ref{E:vertex_condit}) on $B$ rewritten\footnote{ When we need to
remember that $u_j(\cdot)$ also depends on $\lambda$, we will
write it as $u_j(\cdot,\lambda)$.}.

If we now are able to express $Nu_0$ in terms of $u_1(0)$ and
$f_0$, we will essentially separate problems on $\Gamma_0$ and
$\Gamma_1$. This can easily be done due to Lemma \ref{L:DtN}.
Indeed, if $R_0(\lambda)$ is the resolvent of the operator $H_0$
studied in the previous section, then clearly one has
\begin{equation}\label{E:on_gamma0}
    u_0=R_0(\lambda)(\frac{d^2}{dx^2}+\lambda)E(u_1(0))+R_0(\lambda)f_0
\end{equation}
and thus
\begin{equation}\label{E:Robin}
    Nu_0=\Lambda(\lambda)u_1(0)+NR_0(\lambda)f_0=\Lambda(\lambda)u_1(0)+g(\lambda).
\end{equation}
Here, for a given $f_0$ of the considered class,
$g(\lambda)=NR_0(\lambda)f_0$ is a known meromorphic vector
function of $\lambda$ in $\CC$ with singularities only at points
of $\sigma(H_0)$.

Now the problem (\ref{E:spectral_on_gamma1}) can be rewritten as
\begin{equation}\label{E:on_gamma1}
 \begin{cases}
    (-\frac{d^2}{dx^2}-\lambda)u_1=f_1 \mbox{ on }\RR^+\\
    u^\prime_1(0)=\Lambda(\lambda)u_1(0)+g(\lambda).
    \end{cases}
\end{equation}

By the construction, $\Lambda(\lambda)$ is a meromorphic matrix
function in $\CC$ with self-adjoint values along the real axis. We
also observe that the only memory of the compact part of the graph
is confined to the vector-function $g(\lambda)$. We also need to
remember that $u_1$ must belong to $L^2(\RR^+,\CC^n)$.

If we now show that both expressions
$(u_1(\cdot,\lambda),f_1(\cdot))$ and $u_1(0,\lambda)$ continue
analytically through the real axis except a discrete set, then
according to (\ref{E:on_gamma0}) the same will hold for
$(u_0(\cdot,\lambda),f_0(\cdot))$, and thus the Lemma and the main
Theorem will be proven. Hence, we only need to concentrate on the
vector problem (\ref{E:on_gamma1}) on the positive half-axis.

Let us consider the self-adjoint operator $P$ in $L^2(\RR^+)$
naturally corresponding to $-\frac{d^2x}{dx^2}$ with the Neumann
condition at the origin. Let also $r(\lambda)$ be its resolvent.
We sketch below the proof of the following well known limiting
absorption result:
\begin{lemma}\label{L:Neumann}
For any smooth compactly supported function $f$ on $\RR^+$ and any
interval $(a,b)\subset \RR^+$, the inner product $(r(\lambda)f,f)$
as a function of $\lambda$ can be analytically continued through
$(a,b)$ from the upper half-plane $\mbox{Im}\;\; \lambda >0$.
\end{lemma}

Let us chose in the upper half-plane the branch of
$\sqrt{\lambda}$ that has positive imaginary part. The above lemma
then follows immediately from the explicit formula for
$r(\lambda)$:
\begin{equation}\label{E:neumann_explicit}
    (r(\lambda)f)(x) =\frac{i}{2}\int_0^{\infty} \frac{e^{i \sqrt{\lambda} (x+s)} +
e^{i \sqrt{\lambda} |x-s|}}{\sqrt{\lambda}}f(s)\;ds.
\end{equation}
This formula also implies that the value $(r(\lambda)f)(0)$ has
the same analyticity property.


In what follows we will abuse notations using $r(\lambda)$ where
in fact one should use $r(\lambda)\bigotimes I$ (here $I$ is the
unit $n\times n$ matrix).

It is not hard to solve (\ref{E:spectral_on_gamma1}) now. Indeed,
after a simple computation one arrives to the formula for the
solution that one can check directly when $\mbox{Im}\;\;
\sqrt{\lambda}
> 0$:

\begin{equation}\label{E:explicit_gamma1}
  u_1(x,\lambda):= (r(\lambda)f_1)(x) -
i e^{i \sqrt{\lambda} x}A(\lambda)
\end{equation}
where the vector $A(\lambda)$ is:
\begin{equation}
A(\lambda) =\Lambda(\lambda)[\sqrt{\lambda} +
i\Lambda(\lambda)]^{-1}(r(\lambda)f_1)(0)+\frac{g(\lambda)}{\sqrt{\lambda}}
\end{equation}

 Notice that the matrix function $\sqrt{\lambda} +
i\Lambda(\lambda)$ is meromorphic on the Riemann surface of
$\sqrt{\lambda}$. Due to self-adjointness of $\Lambda(\lambda)$,
the values of that function for non-zero real $\lambda$ are
invertible. Hence, the matrix function $\left(\sqrt{\lambda} +
i\Lambda(\lambda)\right)^{-1}$ is meromorphic on the same Riemann
surface.

Now the quantity of interest becomes

\begin{equation}\label{E:analyt_gamma1}
(u_1(\cdot,\lambda),f_1(\cdot))= (r(\lambda)f_1,f_1) -
i(e^{i\sqrt{\lambda} x} A(\lambda),f_1(x)).
\end{equation}

Lemma \ref{L:Neumann} implies the needed analyticity of the first
term in the sum. Since $\mbox{Im}\,\sqrt{\lambda}>0$, according to
the remarks after (\ref{E:neumann_explicit}), $(r(\lambda)f_1)(0)$
is analytic hence $e^{i\sqrt{\lambda} x}A(\lambda)$ is analytic
through $(a,b)$ as well save for a discrete set of $\lambda$. Thus
the final term in the sum can also be analytically continued
through $(a,b)$ as well outside of a discrete set of $\lambda$.

This finishes the proof of Lemma \ref{L:lim_abs}.\qed


Since the space of functions $f$ as above is dense in
$L^2(\Gamma)$, it is well known that (\ref{E:lim_abs}) implies
absence of the singular continuous spectrum (e.g.,
\cite[Proposition 2 and (18) in Section 1.4.5]{Yafaev}, \cite[pp.
136-139 in Section XIII.6]{RS} and thus proves Theorem
\ref{T:main}.

\section{Remarks and acknowledgments}

A procedure similar to the one we use to switch from general Robin
type condition condition to a Neumann condition in
(\ref{E:on_gamma1}) was employed in \cite{Fulling}.

The author would like to thank Prof. Peter Kuchment for his help
and suggestions.

This work was partially supported by the NSF Grants DMS 0296150,
0072248, and 0406022. The author expresses his gratitude to NSF
for this support. The content of this paper does not necessarily
reflect the position or the policy of the NSF and the federal
government, and no official endorsement should be inferred.





\end{document}